\documentclass[fleqn,12pt,twoside]{article}
\usepackage[headings]{espcrc1}

\readRCS
$Id: espcrc1.tex,v 1.2 2004/02/24 11:22:11 spepping Exp $
\usepackage{graphicx}
\newcommand{\beq}{\begin{equation}}
\newcommand{\eeq}{\end{equation}}
\newcommand{\ie}{{\sl i.e.\/}}
\newcommand{\etal}{{\sl et al.\/}}

\newcommand{\tr}{{\mathrm Tr}\,}

\newcommand{\AmS}{{\protect\the\textfont2
  A\kern-.1667em\lower.5ex\hbox{M}\kern-.125emS}}

\title{The critical end point in QCD}

\author{R.\ V.\ Gavai and Sourendu Gupta
        \address{Department of Theoretical Physics, \\
        Tata Institute of Fundamental Research, \\ 
        Homi Bhabha Road, Mumbai 400005, India}%
        \thanks{Talk presented by SG}}
       
\runtitle{QCD end point}
\runauthor{S. Gupta}

\begin{document}

% typeset front matter
\maketitle

\begin{abstract}
In this talk I present the logic behind, and examine the reliability of,
estimates of the critical end point (CEP) of QCD using the Taylor
expansion method.
\end{abstract}

\section{The method and results}

The global structure of the expected phase diagram of two flavour
QCD is the following \cite{expect}.  In the chiral limit of two
flavour QCD, \ie, when $m_\pi=0$, one expects the chiral phase
transition to occur at a critical point at finite temperature, $T$,
and zero baryon chemical potential, $\mu_B$. This critical point
is expected to develop into a critical line in the plane of $T$ and
$\mu_B$ for $m_\pi=0$, and turn into a line of first order transitions
at a tricritical point. When $m_\pi\ne0$, the structure changes
dramatically.  There is no critical line; rather there is a first
order line ending in a critical end point (the CEP) in the Ising
universality class at $T^*(m_\pi)$ and $\mu_B^*(m_\pi)$.  With
varying $m_\pi$ the CEP traces out a ``wing critical line'' ending
at the tricritical point. These considerations are based on symmetry
arguments and universality, and hence are expected to be robust.

Direct verification of this picture involves lattice QCD computations
in the chiral limit and finite chemical potential. Both are technically
unfeasible at this time, the former due to chiral slowing down
(computer time requirements diverge as $m_\pi\to0$) and the latter
due to the sign problem (Monte Carlo is impossible at finite $\mu_B$).
The strategy of \cite{cep} is not to prove or disprove this picture
directly, but, by assuming this picture, to estimate $T^*(m_\pi)$
and $\mu_B^*(m_\pi)$.

This is done by making a Taylor expansion of the pressure around $\mu_B=0$---
\beq
   P(T,\mu_B) \equiv \left(\frac TV\right)\log Z(T,\mu_B) =
     P(T,0) + \sum_n \chi_B^n(T) \frac{\mu_B^n}{n!}.
\label{tep}\eeq
CP symmetry of the problem reduces every odd coefficient to zero, hence
the sum starts with the term in $n=2$. This
induces a series for the quark number susceptibility (QNS), which is
the second derivative of the pressure with respect to $\mu_B$, and the
associated radius of convergence---
\beq
   \frac{\chi(T,\mu_B)}{T^2} 
              \equiv \frac{\partial^2P(T,\mu_B)}{\partial\mu_B^2}
      = \sum_{n=0}^\infty \frac{T^{n-2}\chi_B^{n+2}(T)}{n!}
     \left(\frac{\mu_B}T\right)^n,\qquad
   \rho^{n+1} = \sqrt{\left|
       \frac{n!T^{-2}\chi_B^n(T)}{(n-2)!\chi_B^{n+2}(T)}\right|}.
\label{tec}\eeq
We have used $T$ to convert the series to one in the dimensionless
variable $\mu_B/T$. Since we use this expansion at fixed $T$, there
is complete equivalence with the dimensional form. A notational convention
used here is $\chi(T,\mu_B=0)=\chi_B^2(T)$.

At the CEP this QNS is expected to diverge in the thermodynamic
limit.  Let $\rho^*$ denote the limit of $\rho^n$ as $n\to\infty$,
evvaluated in the thermodynamic limit. Then, somewhere on the circle
of radius $\rho^*(T)$ in the complex plane, we are assured of finding
a singular point.  If this singular point is on the real axis, then
we have found the critical end point.

\begin{figure}[htb]
\begin{minipage}[t]{75mm}
\includegraphics[width=74mm]{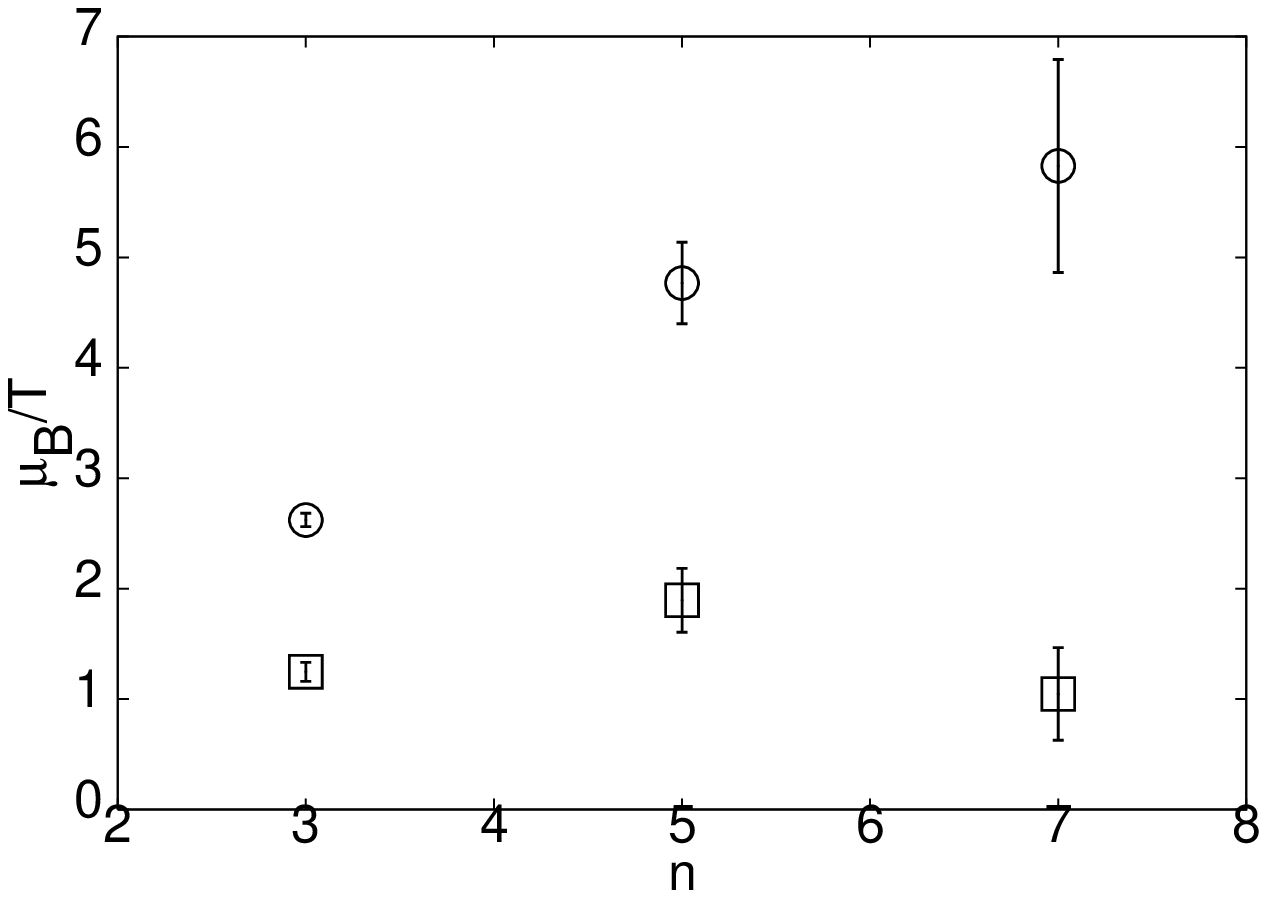}
\caption{$\mu_B^n$ as a function of the order, $n$, for small
($Lm_\pi\simeq3$, circles) and large ($Lm_\pi\simeq9$, boxes) volumes
at $T=0.95T_c$.}
\label{fg.ndep}
\end{minipage}
\hspace{\fill}
\begin{minipage}[t]{75mm}
%\framebox[74mm]{\rule[-26mm]{0mm}{52mm}}
\includegraphics[width=74mm]{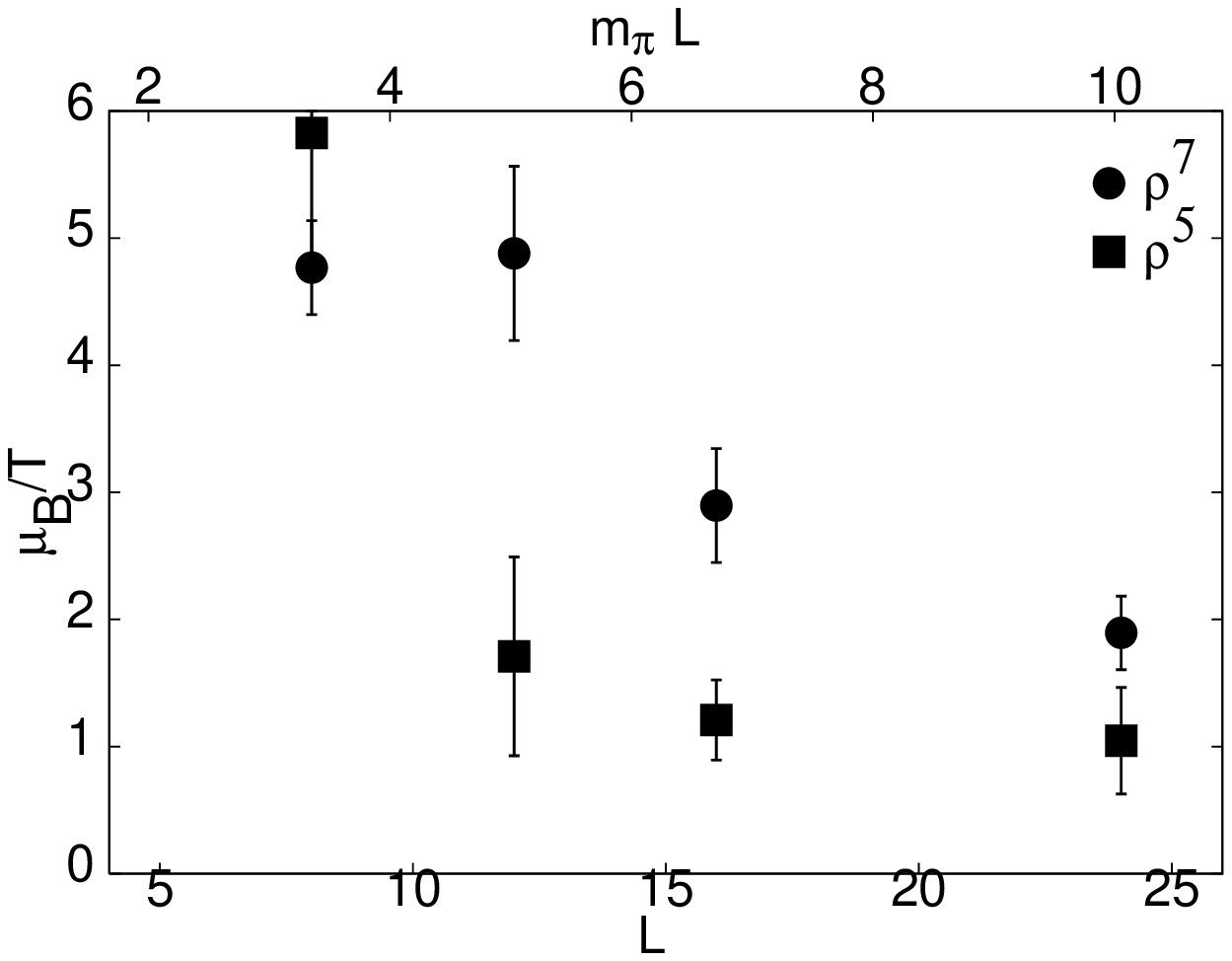}
\caption{$\rho^{5,7}$ as a function of the lattice size, $Lm_\pi$
at $T=0.95T_c$.}
\label{fg.ldep}
\end{minipage}
\end{figure}

One complication is that in a lattice computation one perforce uses
finite spatial volumes--- a cube with sides $L$. On finite volumes
there is no true phase transition, so the limit of $\rho^n$ as
$n\to\infty$ must be infinite. One expects the sequence $\rho^n(T,L)$
to approach some value $\rho^*(T,L)$ for restricted values of $n$,
before climbing to infinity. As $L$ increases, the range of $n$ for
which this stable plateau in $\rho^n(T,L)$ is seen should increase
and the plateau value, $\rho^*(T,L)$, should converge to the
thermodynamic limit $\rho^*(T)$.  In Figure \ref{fg.ndep} we show
that $\rho^n(T,L=3.3/m_\pi)$ increases continually with $n$, whereas
a stable value is seen for $\rho^n(T,L=10/m_\pi)$. In Figure
\ref{fg.ldep} we show that $\rho^{5,7}(T,L)$ cross over from small
to large volume behaviour at $Lm_\pi\ge5$. The extrapolation from
large volumes to the thermodynamic limit can be made using finite
size scaling. In \cite{cep} this is performed assuming that the
CEP is in the Ising universality class.

\begin{figure}[htb]
\begin{minipage}[t]{75mm}
%\framebox[74mm]{\rule[-26mm]{0mm}{52mm}}
\includegraphics[width=74mm]{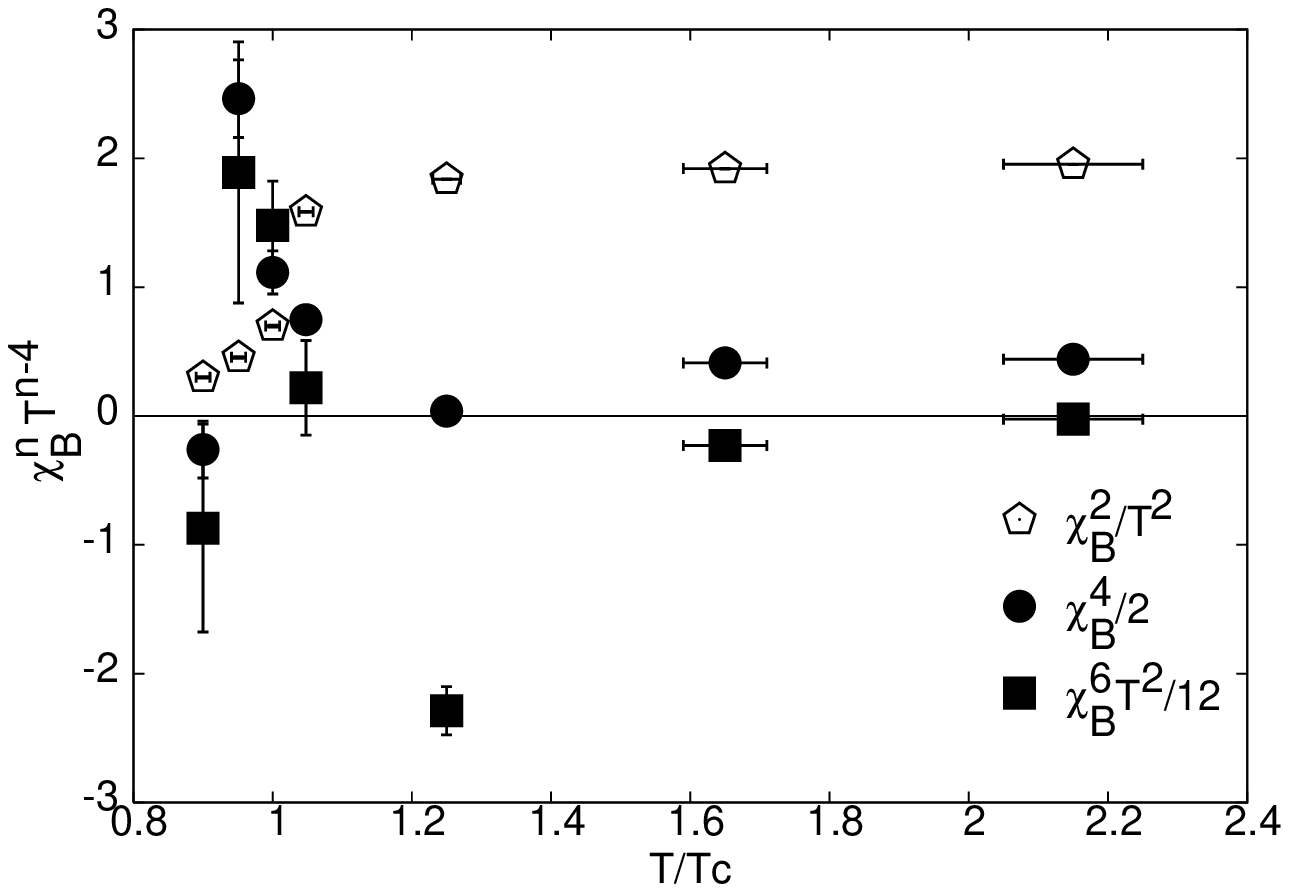}
\caption{The temperature dependence of the successive NLS.}
\label{fg.sign}
\end{minipage}
\hspace{\fill}
\begin{minipage}[t]{75mm}
%\framebox[74mm]{\rule[-26mm]{0mm}{52mm}}
\includegraphics[width=74mm]{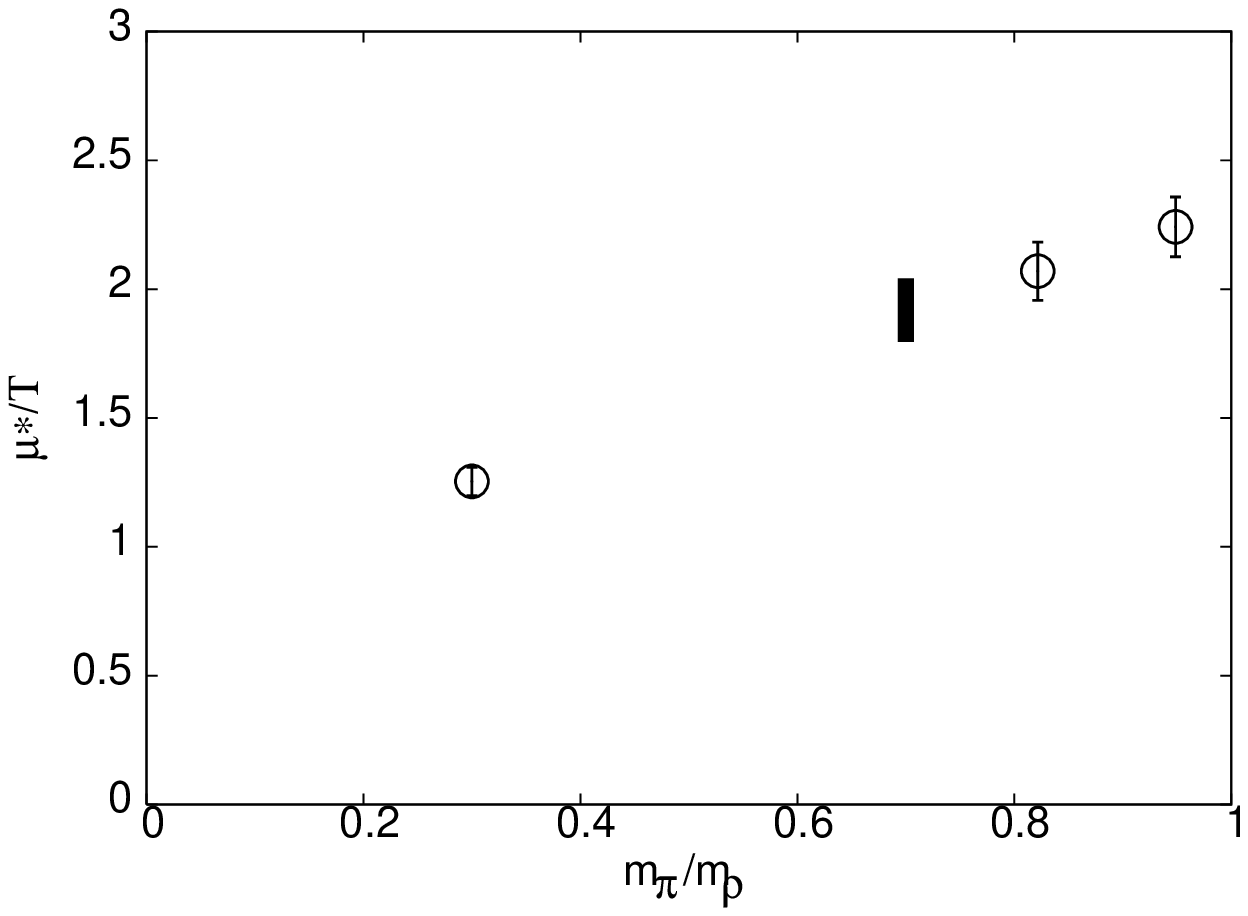}
\caption{Variation of $\rho^3$ with $m_\pi$; the bar is data from \cite{biswa}.}
\label{fg.mpidep}
\end{minipage}
\end{figure}

In \cite{cep} as $T$ was lowered from a very high value, it was
found that the radius of convergence decreased from a very large
value to $\rho^*\approx T$. Oscillations in the sign of the non-linear
susceptibilities (NLS, $\chi_B^n$) indicated that the singularity
was at complex $\rho^*$. On further lowering $T$ it was found that
all the series coefficients have positive sign (Figure \ref{fg.sign}),
showing that the singularity moves to real $\rho^*$, allowing the
identification of $T^*(m_\pi)$ and $\mu_B^*(m_\pi)$.

The technical part of the computation involves the choice of action.
For the gauge fields, we have taken the Wilson action and for the
quarks we have two flavours of staggered quarks. Staggered quarks
at finite chemical potential have many uncertainties at finite
lattice spacing \cite{press,gss}. Since the method of Taylor
expansions uses only simulations at $\mu_B=0$ \cite{cep,biswa},
they may give results which agree with those obtained by reweighting
\cite{fk} or analytic continuation of simulations at imaginary
$\mu_B$ \cite{dp} only in the continuum limit. At present there are
no simulations with Wilson quarks.

The lattice simulations were carried out at lattice spacings $a=1/4T$.
The scale was set by the finite temperature crossover at $T_c$,
giving $m_\rho/T_c=5.4$. The quark mass was chosen such that
$m_\pi/m_\rho=0.31$ (compared to 0.18 in the real world). This agrees
with the simulation parameters of \cite{fk,dp} but not with those
of \cite{biswa} which differ by having $m_\pi/m_\rho=0.7$. 

An exploratory study of the variation of the CEP with $m_\pi/m_\rho$
can be performed in partially quenched QCD, \ie, in QCD where the
sea quark masses are allowed to be different from the valence quark
masses. In such an exercise we found the radius of convergence to
vary by large amounts. We show our results in Figure \ref{fg.mpidep}.
It is interesting that the results of \cite{biswa} are close to
the curve generated by these computations, indicating that the apparent
difference between \cite{cep} and \cite{biswa} could largely be a matter
of change in $m_\pi$.

\section{Two technical points}

A major question in lattice simulations is that of statistics. In
\cite{cep} the measurements were taken on $N_{\mathrm stat}^{\mathrm bare}/(2\tau_{\mathrm
int}+1)=50$--200 statistically independent configurations. Statistical
independence is measured by the integrated autocorrelation time,
$\tau_{\mathrm int}$. Since statistical errors change as
$\sqrt{(2\tau_{\mathrm int}+1)/N_{\mathrm stat}^{\mathrm bare}}$, increase of
statistics by a factor of 4 could easily be offset if $\tau_{\mathrm
int}=1$ instead of zero. Much more important is the statistics used
in the estimator of fermion traces, which we shall discuss next.

\subsection{Evaluating traces and their products}\label{sc.traces}

The Taylor expansion coefficients, the NLS $\chi_B^n$, in eqs.\
(\ref{tep},\ref{tec}) involve derivatives of the quark determinant, which
become traces over quark loops. Traces of very large matrices are
efficiently evaluated by the identity $\tr A=\tr AR$ where $R$ is
a stochastic representation of the identity matrix---
\beq
   R = \frac1{2N_v}\sum_{i=1}^{N_v} \left|r_i\right\rangle
     \left\langle r_i\right| \stackrel{N_v\to\infty}\longrightarrow I,
\eeq
the limit being obtained when $|r_i\rangle$ is one of a set of $N_v$
vectors of complex numbers, each component drawn independently from
a distribution with vanishing mean. Typically one has to evaluate
traces as well as products of traces---
\beq
  a_j = \frac12\sum_{\alpha\beta} \sum_{i=1}^{N_v}
     A_{\alpha\beta} {r^j_i}^*_\alpha {r^j_i}_\beta,\qquad
  p_n = \prod_{j=1}^n a_j,\qquad
  \overline{p_n} = \prod_j\tr A_j = \prod_j \left[\frac1{N_v}\sum_{i=1}^{N_v}
       a_j\right] = \prod_j \overline{a_j},
\eeq
where the factorisation of the average (denoted by a bar over the symbol)
of $p_n$ occurs when the different sets of random vectors $r^j$ are
independent. Here $\alpha$ and $\beta$ are matrix indices. If the
random vectors are drawn from a Gaussian distribution, then $a_j$
is clearly a Gaussian distributed variable with mean $\overline{a_j}=\tr
A_j$ and variance $\sigma_j^2\propto1/N_v$. Hence the product of
traces is a product of independent Gaussian random variables.

A product of Gaussian random variables is not Gaussian. To prove
this, consider $\delta_j=a_j-\overline{a_j}$, which is Gaussian
distributed with zero mean, so $\langle\delta_j^2\rangle=\sigma_j^2$
and $\langle\delta_j^4\rangle=3\sigma_j^4$, so that $[\delta_j^4]\equiv
\langle\delta_j^4\rangle-3\langle\delta_j^2\rangle=0$, as must be
true for a Gaussian. In that case, however, the product
$d=\prod_{j=1}^n\delta_j$ has the property that
\beq
   \langle d^2\rangle=\prod_{j=1}^n\langle\delta_j^2\rangle=
       \prod_{j=1}^n\sigma_j^2,\qquad
   \langle d^4\rangle=\prod_{j=1}^n\langle\delta_j^4\rangle=
       3^n\prod_{j=1}^n\sigma_j^4,\qquad
   [d^4]=(3^n-3)\prod_{j=1}^n\sigma_j^4\ne0,
\eeq
where the factorisation of the expectation values follows from the
independence of the $\delta_j$'s. Thus products of Gaussians are
highly non-Gaussian. 

Since the central limit theorem applies, estimators of $\overline
d$ are still Gaussian distributed.  It would seem that the fourth
cumulant varies as $(3^n-3)/N_v^{2n}$, and therefore the 
Gaussian is approached rapidly.  In actuality the situation
is worse, as one can check by reconstructing the distribution of
$p_n=\prod_j (\overline{a_j}+\delta_j)$.  We find that $[p_n^4]$
is dominated by the slowest falling term, which varies as $n^2/N_v^2$,
rather than the most rapidly falling term $(3/N_v^2)^n$. It was
found that $N_v\simeq30$ are needed to control errors for $n=2$
\cite{press}. Consistent with this, $N_v\simeq500$ are needed to
control errors for $n=8$ \cite{cep}.  Conversely, using $N_v=100$
for $n=6$ \cite{biswa} is equivalent to working with $N_v\simeq 10$
for $n=2$.

\subsection{Critical divergence}\label{sc.ising}

\begin{figure}[htb]
\begin{minipage}[t]{75mm}
%\framebox[74mm]{\rule[-26mm]{0mm}{52mm}}
\includegraphics[width=74mm]{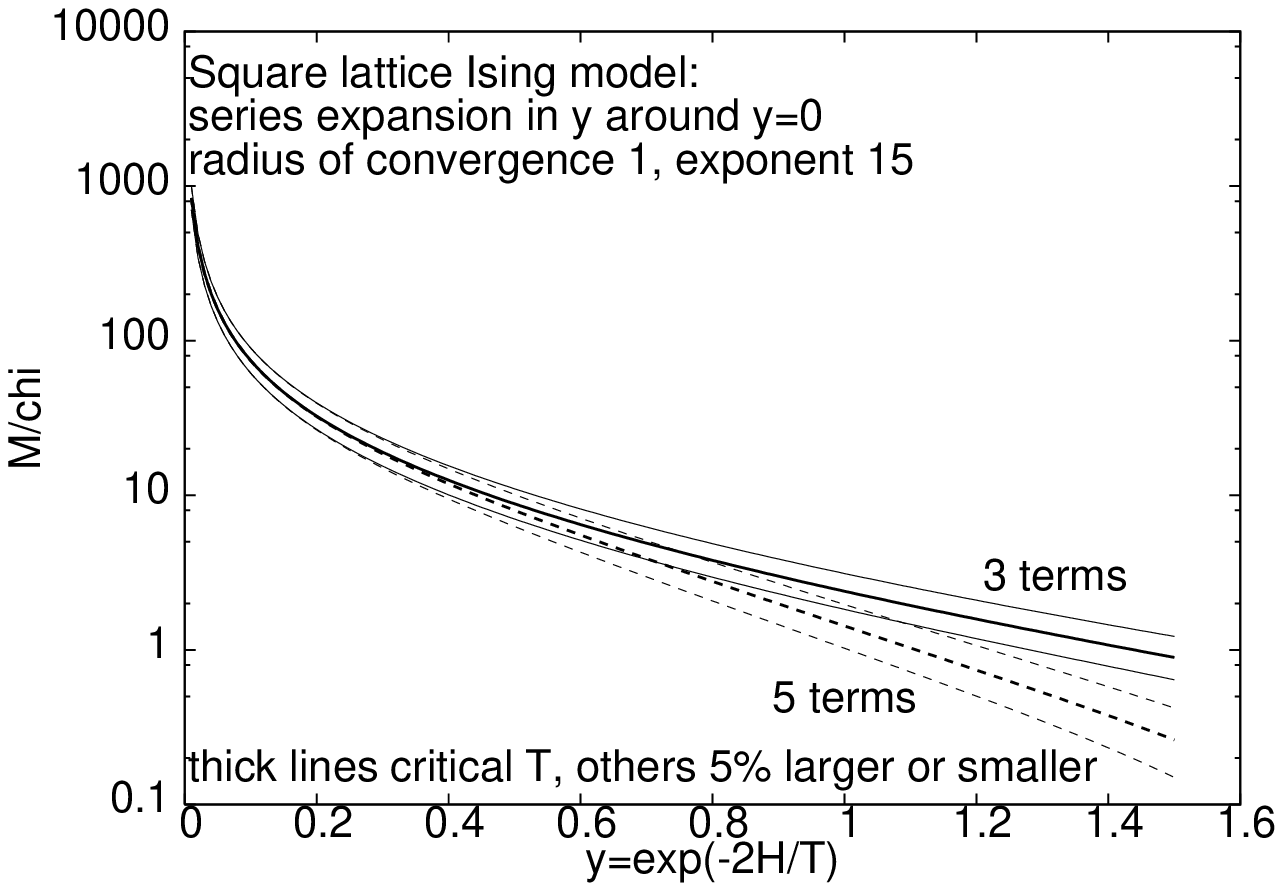}
\caption{Partial series for the ratio $m/\chi$. There is no hint of
  critical behaviour at $y=1$.}
\label{fg.mag}
\end{minipage}
\hspace{\fill}
\begin{minipage}[t]{75mm}
%\framebox[74mm]{\rule[-26mm]{0mm}{52mm}}
\includegraphics[width=74mm]{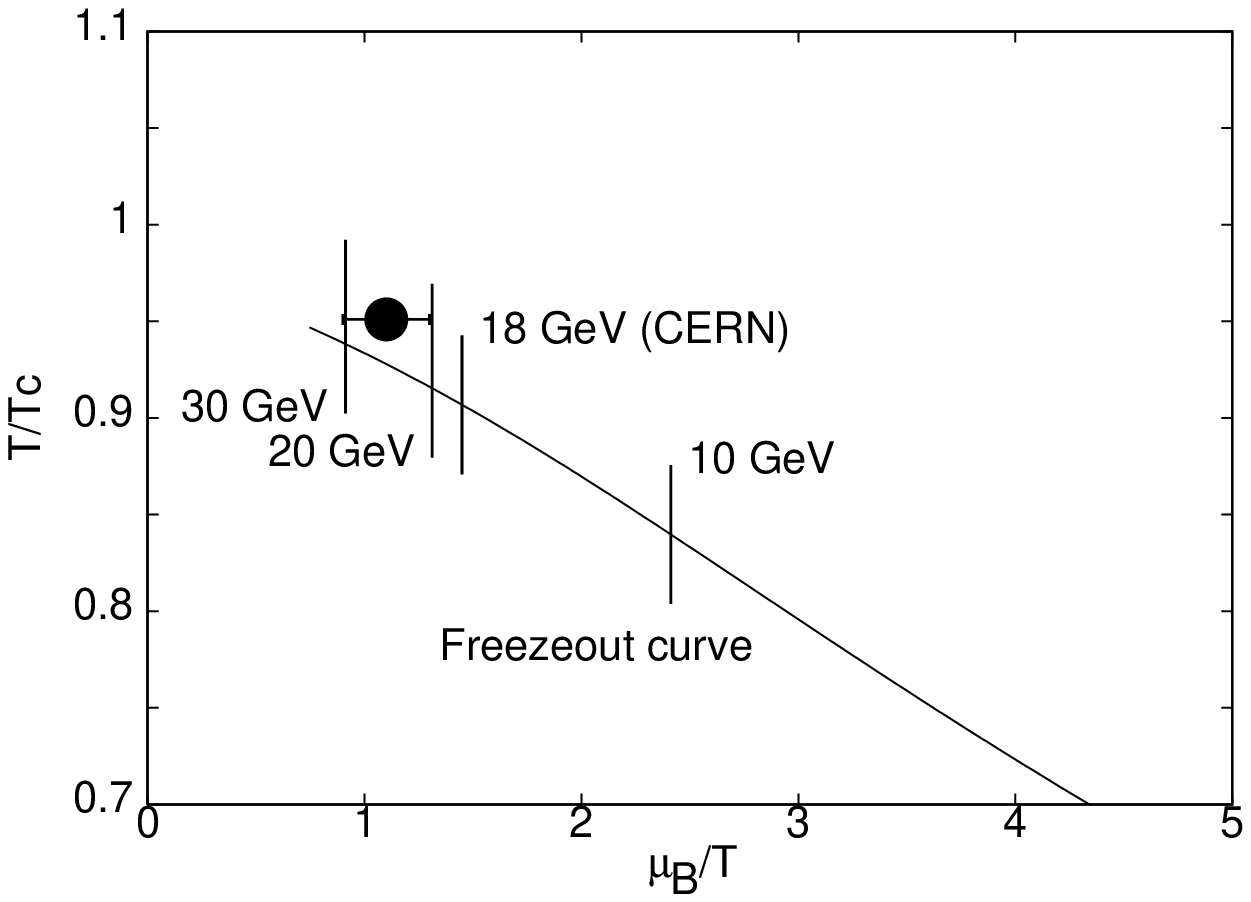}
\caption{A comparison of the CEP obtained in \cite{cep}
with the freezeout curve in heavy-ion collisions \cite{cley}.}
\label{fg.freeze}
\end{minipage}
\end{figure}

Since we analyze the series expansion of eq.\ (\ref{tec}) to estimate
the critical end point shouldn't one be able to use the series to
see critical divergence? In fact, the lack of such a divergence has
been used as an argument for the absence of the critical end point
in \cite{biswa}. However, it has been known since the 1960's that
the partial sum of the series expansion is generically perfectly
smooth across the critical point as determined by the radius of
convergence upto the same order.

As an illustration I show results for the series expansion of the
magnetic susceptibility, $\chi$, of the two dimensional Ising model
on a square lattice. The expansion is made in $y=\exp(-2H/T)$ around
the point $y=0$, where $H$ is the magnetic field and $T$ the
temperature. This series has a radius of convergence unity. The
exact ratio $m/\chi$ (where $m$ is the magnetization) vanishes in
the ordered phase. However, as shown in Figure \ref{fg.mag}, the
ratio is perfectly smooth across $y=0$.

One consequence of this realization is that the agreement of a
series expansion of the pressure summed to low orders with any
effective theory, for example, a resonance gas model, cannot be
used to exclude the possibility of a critical end point.

\section{Summary}

There are four measures of reliability of lattice simulations of QCD---
\begin{enumerate}
\item the lattice spacing, $a$, which must be taken to zero, but
   in current computations is $1/4T$ \cite{cep,fk,dp,biswa,fk2}.
   The next generation of computations will
   reduce the lattice spacing by 33\%.
\item the quark mass, which has to be tuned by fixing the pion mass
   or the ratio $m_\pi/m_\rho$. The physical value of this ratio is
   0.18. In \cite{cep,fk,dp} this was taken be 0.31, in \cite{biswa}
   it is 0.7 and in \cite{fk2} it is the physical value.
\item the spatial size of the lattice, $L$, which must be taken to
   infinity through the process known as finite size scaling in order
   to obtain reliable results for thermodynamics. It was found that
   $Lm_\pi<5$ gave rise to tremendous finite size distortions \cite{cep}.
   Only \cite{cep,biswa} use lattices larger than this.
\item the statistical precision, $N_{\mathrm stat}^{\mathrm true} = \sqrt
   {N_{\mathrm stat}^{\mathrm bare} / (2 \tau_{\mathrm int} + 1)}$, where
   $N_{\mathrm stat}^{\mathrm bare}$ is the number of configurations
   generated and $\tau_{\mathrm int}$ is a measure of statistical
   independence called the integrated autocorrelation time.  When
   $\tau_{\mathrm int}=0$, the naive statistics actually represents the
   true statistics, but even the small value $\tau_{\mathrm int}=1$, causes
   the true statistics to be three times smaller than $N_{\mathrm stat}^{\mathrm bare}$. In
   \cite{cep} $N_{\mathrm stat}^{\mathrm true}=50$--200 at all $T$, whereas no other
   simulation reports measurements of $\tau_{\mathrm int}$.
\end{enumerate}

\noindent
In addition, any computation of the expectation of products of fermion
traces contains another measure of reliability. This is the number of
vectors, $N_v$, used in evaluating the traces. Its crucial role is
explained in Section \ref{sc.traces}.

The method of Taylor expansions can give reliable answers to questions
about the QCD phase diagram once all these technical matters are under
control. The critical end point can be determined in two steps---
\begin{enumerate}
\item Generate the series coefficients for the QNS in eq.\ (\ref{tec})
   and evaluate the radius of convergence of the series.
\item When all the series coefficients have the same sign, the radius
   of convergence implies a singularity at real $\mu_B$, which is the
   location of the CEP.
\end{enumerate}

\noindent
This works although the partially summed series for the pressure (eq.\
\ref{tep}) or the QNS (eq.\ \ref{tec}) are not expected to give the
physical values of these quantities, as we show through examples in
section \ref{sc.ising}.  As a result, comparison of models, such as
the resonance gas model, with the partial series sums is blind to the
physics of the critical end point.

The CEP of QCD can be obtained from the results presented in \cite{cep},
with $Lm_\pi>5$, pion mass within 50\% of the physical pion mass, and with
good statistical control. The result is shown in Figure \ref{fg.freeze}
superposed on the freezeout curve obtained in \cite{cley}. Computations
at smaller lattice spacings would be very useful in checking the stability
of this result.

\end{document}